# Hydrogen-Atom Electronic Basis Sets for Multicomponent Quantum Chemistry

Irina Samsonova, Gabrielle B. Tucker, Naresh Alaal, Kurt R. Brorsen*

Department of Chemistry, University of Missouri, Columbia, Missouri 65203, USA

*email: brorsenk@missouri.edu

**Abstract**

Multicomponent methods are a conceptually simple way to include nuclear quantum effects into quantum chemistry calculations. In multicomponent methods, the electronic molecular orbitals are described using the linear combination of atomic orbitals approximation. This requires the selection of a one-particle electronic basis set which, in practice, is commonly a correlation-consistent basis set. In multicomponent method studies, it has been demonstrated that large electronic basis sets are required for quantum hydrogen nuclei to accurately describe electron-nuclear correlation. However, as we show in this study, much of the need for large electronic basis sets is due to the correlation-consistent electronic basis sets not being optimized to describe nuclear properties and electron-nuclear correlation. Herein, we introduce a series of correlation-consistent electronic basis sets for hydrogen atoms called cc-pV$n$Z-mc with additional basis functions optimized to reproduce multicomponent density functional theory protonic densities. These new electronic basis sets are shown to yield better protonic densities with fewer electronic basis functions than the standard correlation-consistent basis sets and reproduce other protonic properties such as proton affinities and protonic excitation energies, even though they were not optimized for these purposes. The cc-pV$n$Z-mc basis sets should enable multicomponent many-body calculations on larger systems due to the improved computational efficiency they provide for a given level of accuracy.

## I. Introduction

Multicomponent methods include nuclear quantum effects directly into quantum chemistry calculations by not invoking the Born-Oppenheimer approximation for select or all nuclei in a system.[1-3] In multicomponent methods, the electrons and quantum nuclei occupy molecular orbitals that are described using the linear combination of atomic orbitals method. As in standard (or single-component) quantum chemistry, these atomic orbitals are typically Gaussian functions. An atomic-orbital basis set is defined as a particular set of these functions. In practice, multicomponent methods decompose the atomic-orbital basis set into separate electronic and nuclear basis sets. Single-component electronic basis sets are normally used without modification. Historically, it has been common for multicomponent methods to use an even-tempered[4] nuclear basis set for the nuclear molecular orbitals.[5-8] More recently, the PB family of nuclear basis sets,[9] which are uncontracted nuclear basis sets with exponents fit to reproduce protonic densities, proton affinities, and protonic excitations energies, has seen wide adoption due to its increased computational efficiency and numerical stability in comparison to even-tempered nuclear basis sets.

Multicomponent-method studies have consistently shown that accurately describing electron-nuclear correlation requires the use of large electronic basis sets for the nucleus or nuclei being treated quantum mechanically.[10-16] Many-body multicomponent methods such as orbital-optimized second-order Møller-Plesset perturbation theory[14, 17] or coupled-cluster theory[10, 12, 16, 18] normally require electronic basis sets of at least quadruple-zeta quality for the quantum nucleus to obtain accurate energies or protonic densities. Protonic excited-state calculations using multicomponent time-dependent density-functional theory[19] require electronic basis sets of up to sextuple quality on the quantum proton.[11] Due to requiring large electronic basis sets, and the fact that multicomponent methods have the same computational scaling with respect to system size as their single-component counterparts, multicomponent calculations can be computationally expensive. In order to mitigate this computational expense, multicomponent calculations are often performed with a mixed electronic basis set in which the quantum nucleus (or nuclei) uses a basis set with a higher zeta level than the atoms of the system that are treated classically.[10-13, 19-22]

However, the usage of mixed electronic basis sets has been questioned. A recent multicomponent coupled cluster with single and double excitations and perturbative triple excitations (CCSD(T)) study[16] showed that the good performance of multicomponent CCSD proton affinities[10] was likely due to a cancelation of errors that arises when using a mixed basis set of aug-cc-pVQZ for the quantum proton and aug-cc-pVTZ for all other atoms. When the aug-cc-pVQZ basis set was used for all atoms, the mean absolute error (MAE) of multicomponent CCSD proton affinities for a test set of 12 small molecules increased by a factor of four relative to the mixed basis-set results. The cancelation of errors in multicomponent CCSD with a mixed electronic basis set was hypothesized to arise from a decrease in the basis set error for a subset of the system in the region of the quantum proton, which cancels the incomplete description of the electron-nuclear correlation. Further evidence for this claim was demonstrated by the better performance of multicomponent CCSD(T) for the calculation of proton affinities when using the aug-cc-pVQZ electronic basis set for all atoms, which had an MAE of 0.05 eV, in comparison to calculations performed using a mixed electronic basis set that had a larger MAE of 0.09 eV.

To understand how to improve the electronic basis sets used in multicomponent calculations, it is useful to review the development of multicomponent nuclear basis sets. In principle, it is possible for nuclear basis sets to be identical to electronic basis sets, as both consist of Gaussian functions. However, electronic basis sets are predominantly designed to provide a flexible description of the electronic valence region of a molecule with additional flexibility as needed in other regions of space. In multicomponent methods, the quantum nuclei occupy delocalized nuclear orbitals, but are still relatively well localized compared to electrons and are therefore localized in a different region of space than the electronic valence region of a molecule. For this reason, basis sets with exponents optimized for the nuclear orbitals were introduced.

During the initial stages of the development of multicomponent methods, it was common to use a single s-type Gaussian function for the nuclear basis set and variationally optimize the exponent during the multicomponent Hartree-Fock calculation.[1, 23-25] However, using a single s-type function prevents the nuclear orbitals from polarizing, which results in reduced accuracy. Furthermore, extending the approach to the optimization of exponents for multiple Gaussian functions for each multicomponent calculation is computationally impractical. Therefore, nuclear basis sets were introduced by variationally optimizing the exponents of a collection of Gaussian functions for a set of molecules using multicomponent Hartree-Fock.[2] The nuclear basis sets were then able to be used for multicomponent calculations on other systems. While these basis sets are more accurate than a basis composed of a single s-type orbital, their small size (2s2p2d was the largest nuclear basis set introduced in the original study), leads to insufficient flexibility, especially for multicomponent calculations that accurately include electron-nuclear correlation.

The need for additional flexibility in the nuclear basis set led to the use of even-tempered[4] nuclear basis sets. These are uncontracted basis sets where the set of exponents of a given shell $\{\xi_i\}$ are a logarithmic series such that $\xi_i = \alpha\beta^i, i = 1, 2, 3, \ldots N$, with $\alpha$ and $\beta$ being parameters that define the basis set.[5-8] The most commonly used even-tempered nuclear basis set is an 8s8p8d nuclear basis set with $\alpha = 2$ and $\beta = \sqrt{2}$.

Even-tempered nuclear basis sets are accurate for the calculation of nuclear properties, but this accuracy comes at a cost because it is achieved by using an essentially dense set of functions. Therefore, even-tempered basis sets are highly linearly dependent, which can make it difficult to converge multicomponent Hartree-Fock and density-functional theory (DFT) calculations. For example, the overlap matrix of the cartesian 8s8p8d nuclear basis set with $\alpha = 2$ and $\beta = \sqrt{2}$ has 16 near-zero eigenvalues (less than $10^{-5}$) out of 80 total basis functions. Even for the same 8s8p8d nuclear basis set with spherical basis functions, the linear dependencies remain with 10 eigenvalues of the overlap matrix being less than $10^{-5}$ out of a total of 72 basis functions.

Additionally, these nuclear basis sets result in high computational cost due to the number of basis functions they contain. For example, a medium-sized spherical 8s8p8d8f nuclear basis set has 128 basis functions for a single nucleus. The high computational scaling of many-body multicomponent methods can make calculations with even-tempered basis sets difficult, especially when multiple nuclei are treated quantum mechanically.

The most common quantum nuclei in multicomponent methods are hydrogen nuclei, (i.e., protons). To address the aforementioned basis set issues for quantum protons, the PB family of nuclear basis sets was introduced in 2020.[9] All PB basis sets are uncontracted nuclear basis sets. Sets of exponents for each basis were found by a Gaussian process regression[26-27] that minimized a fitness function of protonic densities, proton affinities, and protonic excitation energies. The PB basis sets mitigate the issues in the previous two paragraphs. As an example, the PB4D protonic basis set is of similar accuracy to the 8s8p8d even-tempered basis set for protonic properties while reducing the number of protonic basis functions by a factor of three, but with no eigenvalues of the overlap matrix that are less than $10^{-5}$. The PB protonic basis sets have been rapidly adopted by the multicomponent community and have allowed for the study of much larger systems using multicomponent many-body methods.

Multicomponent systems require both nuclear (protonic) and electronic basis sets, and it has been previously shown that accurate description of electron-nuclear correlation requires a large electronic basis set for the electronic basis functions associated with the quantum nuclei. This significantly increases the computational cost, especially when calculating the electron-electron correlation. Additionally, this larger basis set must be used with all atoms in order to avoid the previously mentioned issues that can arise with mixed basis sets. To date, there has been little effort devoted to reducing the need for large electronic basis sets in multicomponent calculations. Previous suggestions in the literature include using an F12 correction[28] to satisfy the electronic-nuclear cusp condition or the extrapolation of multicomponent calculations using different electronic basis sets.[18] Both ideas have been widely used in single-component quantum chemistry.[29-33] While both are useful possibilities, F12 many-body methods can be complicated and no multicomponent F12 method has been implemented. Extrapolation methods can still require running calculations with basis sets of quadruple or quintuple zeta-level quality, which could be computationally expensive. In this study, we take a different approach by augmenting electronic hydrogen atom basis sets with basis functions and optimizing those functions to reproduce protonic multicomponent DFT densities.

Electronic basis sets designed for calculations on specific types of systems or properties, such as anions[34-35] or nuclear magnetic resonance coupling constants[36-37] have been used extensively in single-component quantum chemistry. To better understand how to develop new electronic basis sets for multicomponent methods, reviewing the development of protonic basis sets with exponents fit to reproduce protonic properties offers some guidance, but additional complications arise in the electronic case. For example, take the standard cc-pV$n$Z hydrogen-atom basis sets[34-35] used often in single-component calculations, which are the basis sets that are augmented in this study. The cc-pV$n$Z electronic basis sets were designed to converge smoothly to the complete basis set limit with increasing zeta level when using correlated single-component methods. If new electronic basis sets are fit to reproduce multicomponent energetic quantities such as proton affinities, they could achieve these results by including more electron-electron correlation energy, but this would destroy the systematic convergence behavior in the electronic correlation energy. This effect would be similar to the hypothesis as to why multicomponent CCSD mixed electronic basis set proton affinities agree well with experimental values. Ideally, we would like to add the new electronic functions with no change to the electron-electron correlation energy, but this is clearly not strictly possible.

We still wish to minimize the change in the electron-electron correlation energy when the new electronic basis functions are included in the cc-pV$n$Z basis sets. Therefore, rather than optimize a fitness function of protonic densities, proton affinities, and protonic excitation energies like in the fitting procedure for the PB family of protonic basis sets,[9] we optimize a fitness function that consists only of protonic densities. With the additional electronic basis functions obtained from this fitting procedure, we introduce new electronic basis sets, denoted cc-pV$n$Z-mc, designed for multicomponent calculations. As shown in this study, for a given zeta level, cc-pV$n$Z-mc protonic density errors are lower than values calculated using standard cc-pV$n$Z basis sets two zeta levels higher. Even though the new basis sets were designed to reproduce protonic densities at the multicomponent DFT level, we show that they are transferable to other multicomponent methods and protonic properties such as multicomponent CCSD(T) proton affinities and multicomponent heat-bath configuration interaction (HCI) protonic excitation energies. Due to the increased accuracy for a given number of electronic basis functions, these basis sets should facilitate multicomponent calculations on larger systems and allow for longer time scales in multicomponent *ab initio* molecular dynamics.[38-40]

## II. Methods

In this study, we introduce new correlation-consistent electronic basis sets for hydrogen atoms that include additional electronic basis functions with exponents fit to reproduce multicomponent DFT protonic density calculations with a large electronic basis set. These basis sets are denoted cc-pV$n$Z-mc, where $n$ is the zeta level of the basis set. We introduce new electronic basis sets at the DZ, TZ, and QZ levels. The number and shell types of the additional electronic basis functions at each of these zeta levels are 2s1p, 3s2p1d, and 3s3p2d1f respectively. All of the new electronic basis functions are uncontracted.

To generate reference protonic densities for the fitting procedure, multicomponent DFT calculations were performed on the HCN and FHF$^-$ molecules with the hydrogen nucleus of each molecule treated quantum mechanically. These two systems were previously used in the fitness function optimization of the protonic basis set for the PB nuclear basis sets.[9] All reference calculations used the PB5F protonic basis set and the cc-pV6Z electronic basis set. The cc-pV6Z electronic basis set includes electronic basis functions of *i* angular momentum for non-hydrogenic atoms. Because GAMESS,[41] which was used to perform all of the multicomponent DFT calculations, lacks the capability to compute two-particle atomic-orbital integrals using *i* basis functions, these functions were excluded from the reference calculations. As the electronic basis functions that we add in this study are included only on hydrogen atoms, it is likely that the lack of *i* functions on non-hydrogen atoms has only a small effect on the fitting procedure.

Though the Fourier-grid Hamiltonian (FGH) method [42-43] is the standard reference for multicomponent methods, in this study, we use multicomponent DFT protonic densities as a benchmark. This is because while the FGH method with a well-chosen grid is at the complete nuclear or protonic basis set limit, it also includes all electron-nuclear correlation for the quantized particle for a given electronic basis set. In this study, we want to isolate the effects of the incomplete nuclear and electronic basis set from the incomplete multicomponent DFT description of electron-nuclear correlation, which is difficult using the FGH method.[21] Therefore, this study seeks to reproduce multicomponent DFT protonic densities with the cc-pV6Z electronic basis.

The B3LYP electronic exchange-correlation functional[44-46] and epc17-1 electron-proton correlation functional[7] were used for all reference protonic density calculations. As has been previously shown, the choice of electronic exchange-correlation functional is less important than the choice of electron-proton correlation functional to obtain accurate protonic properties.[47] The geometries for the calculations were obtained from a multicomponent DFT geometry optimization using the same parameters as the reference density calculations. For all HCN and FHF⁻ calculations, the optimized geometry was chosen such that the position of the hydrogen electronic and nuclear basis functions was located at the origin, with the classical nuclei of the molecule aligned on the z-axis. For each of the molecules, a reference protonic density was generated on a three-dimensional grid from -0.6 to 0.6 Å with 33 evenly spaced points in each dimension.

To determine the exponents of the additional electronic basis functions, an iterative procedure was performed for each basis set. For each shell, optimal exponents were found using a grid-based search. For shells with three basis functions, a further constraint was applied such that the exponents were even-tempered, which reduced the three-dimensional grid search to a two-dimensional grid search. For each set of exponents generated during the fitting procedure, protonic densities were computed on the three-dimensional grid from multicomponent DFT calculations on the HCN and FHF$^-$ molecules. These calculations were performed identically to the reference density calculations, including the position of the classical nuclei and quantum nuclei basis functions, with the exception of using a different electronic basis set. The root-mean-squared error (RMSE) in the protonic density on the 3D grid was computed for each molecule relative to the reference protonic density and the average of the two RMSEs was calculated to determine which set of exponents from the grid search was the most accurate. The procedure was repeated to find optimal exponents for the next shell type. After the exponents of each shell had been optimized, the entire procedure was repeated starting with the s-shell with a finer grid centered on the previous optimal exponent(s) or even-tempered values. The iterations were halted when the change in the average of the protonic density RMSEs at least fell below $10^{-7}$. A schematic of the optimization procedure is presented in Figure 1.

The exponents of the basis functions added to the cc-pV*n*Z electronic basis sets to construct the cc-pV*n*Z-mc electronic basis sets are shown in Table 1. Spatially, atomic orbitals with these exponents are located in the electronic core region of most atoms. These exponents are similar in magnitude to those of the nuclear exponents in the PB family of protonic basis sets,[9] so the new basis functions give the electronic basis set more flexibility in the regions of space with a large protonic density.

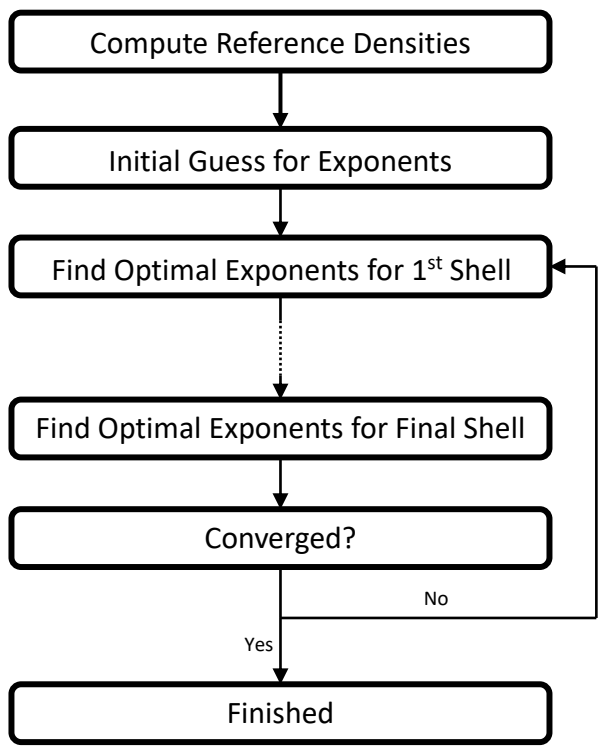

**Figure 1:** Scheme of the procedure used to determine the optimal exponents for the additional basis functions.

| Basis Set | DZ | TZ | QZ |
|---|---|---|---|
| Definition | 2s1p | 3s2p1d | 3s3p2d1f |
| s-type | 2.32727 | 1.50000 | 5.76923 |
|  | 11.03922 | 11.38180 | 6.64506 |
|  |  | 86.36364 | 7.65385 |
| p-type | 2.31579 | 0.65385 | 12.00000 |
|  |  | 2.65357 | 18.30511 |
|  |  |  | 27.92308 |
| d-type |  | 8.55789 | 4.30769 |
|  |  |  | 13.28967 |
| f-type |  |  | 2.89474 |

**Table 1:** Shell types with exponents for the additional basis functions in the cc-pV$n$Z-mc electronic basis sets.

**III. Results and Discussion**

The cc-pV$n$Z-mc electronic basis sets are benchmarked against multicomponent calculations of protonic densities, proton affinities, and protonic excited states. All three properties are standard benchmarks of new multicomponent methods.[3] For the proton affinity calculations, aug-cc-pV$n$Z-mc electronic basis sets are used. The aug-cc-pV$n$Z-mc basis sets are constructed by adding the diffuse functions of the aug-cc-pV$n$Z electronic basis sets to the cc-pV$n$Z-mc electronic basis sets.

The first benchmark is the multicomponent DFT protonic densities of FHF$^-$ and HCN. These calculations were performed identically to the protonic density calculations of Section II that found the optimal exponent values of the additional electronic basis functions. Because the exponents of the additional electronic basis functions were fit using these two systems, the protonic density of the HNC molecule was also calculated, with the hydrogen nuclei treated quantum mechanically. The results from these calculations are presented in Table 2.

As Table 2 demonstrates, multicomponent DFT protonic densities calculated with the cc-pV$n$Z-mc electronic basis sets are more accurate than calculations with the cc-pV$n$Z electronic basis sets relative to the reference densities. For all systems, the accuracy of the cc-pVDZ-mc densities is comparable to cc-pVQZ densities and the accuracy of the cc-pVTZ-mc densities is comparable to cc-pV5Z densities. From these results, it appears that for the calculation of protonic densities, calculations performed with the cc-pVnZ-mc electronic basis sets are as accurate as those performed with cc-pVnZ electronic basis sets that are two zeta levels higher. This is also true for the HNC molecule whose protonic density was not considered in the exponent fitting procedure. However, the HNC molecule does have higher RMSEs than the HCN molecule for the cc-pV$n$Z-mc electronic basis sets.

|  | cc-pVnZ | | | | cc-pVnZ-mc | | |
| --- | --- | --- | --- | --- | --- | --- | --- |
| Molecule | DZ | TZ | QZ | 5Z | DZ | TZ | QZ |
| FHF- | 3.71E-04 | 2.95E-04 | 1.84E-04 | 7.99E-05 | 1.33E-04 | 4.18E-05 | 2.13E-05 |
| HCN | 4.03E-04 | 5.66E-04 | 2.43E-04 | 1.14E-04 | 2.42E-05 | 1.27E-05 | 1.82E-05 |
| HNC | 8.14E-04 | 5.33E-04 | 2.48E-04 | 9.20E-05 | 1.64E-04 | 6.48E-05 | 6.05E-05 |

**Table 2:** Protonic density RMSEs relative to the reference density for different electronic basis sets. The RMSEs are computed over the set of 3D grid points identical to that used in the fitting procedure in Section II. All values are in atomic units.

With the development of excited-state and *ab initio* molecular dynamics multicomponent methods, accurately calculating energetic quantities has become essential in the field of multicomponent methods. Therefore, aug-cc-pVnZ-mc proton affinities are examined next. We emphasize that in the optimization of the exponents of the additional aug-cc-pVnZ-mc basis functions, no energetic data was included.

For multicomponent methods with a single quantum proton, the proton affinity of species A, $PA(A)$, can be computed as

$$PA(A) = E_A - E_{AH^+} + \frac{5}{2}RT, \qquad (1)$$

where $E_A$ is the energy obtained from a single-component geometry optimization and $E_{AH^+}$ is the energy of a multicomponent calculation, with the most acidic hydrogen treated quantum mechanically at a level of theory analogous to the single-component calculation for $E_A$. The geometry for the multicomponent calculation on the AH$^+$ system is obtained from a single-component geometry optimization of AH$^+$. All DFT calculations used the B3LYP electronic exchange-correlation functional, and all multicomponent DFT calculations used the epc17-2 electron-proton correlation functional[48] and the PB4F-1 protonic basis set. For calculations that used the aug-cc-pVnZ-mc electronic basis sets, the aug-cc-pVnZ-mc electronic basis set was used for both quantum and classical hydrogen atoms. Further discussion about the derivation and assumptions of Eq. 1 can be found in the literature.[3, 10]

Multicomponent DFT proton affinities for a test set of 12 small molecules are shown in Table 3. From these results, we see that while the aug-cc-pV$n$Z-mc electronic basis sets have different proton affinities for individual molecules relative to the aug-cc-pV$n$Z electronic basis sets, their overall performance is similar, and the MAE of basis sets with different zeta levels is nearly identical. As is well established in single-component computational chemistry, DFT is normally less sensitive for energy calculations to the choice of the electronic basis set than many-body methods,[49-51] which likely explains the similar performance of all of the basis sets in the calculation of proton affinities.

| Molecule | Experiment | aug-cc-pV$n$Z | | | aug-cc-pV$n$Z-mc | | |
|---|---|---|---|---|---|---|---|
|  |  | DZ | TZ | QZ | DZ | TZ | QZ |
| $CN^-$ | 15.31 | -0.24 | -0.20 | -0.18 | -0.19 | -0.15 | -0.15 |
| $NO_2^-$ | 14.75 | -0.14 | -0.09 | -0.07 | -0.08 | -0.04 | -0.04 |
| $NH_3$ | 8.85 | 0.01 | 0.02 | 0.03 | 0.05 | 0.06 | 0.06 |
| $HCOO^-$ | 14.97 | -0.13 | -0.10 | -0.08 | -0.07 | -0.05 | -0.05 |
| $HO^-$ | 16.95 | -0.14 | -0.11 | -0.09 | -0.08 | -0.07 | -0.07 |
| $HS^-$ | 15.31 | -0.23 | -0.19 | -0.17 | -0.18 | -0.14 | -0.14 |
| $H_2O$ | 7.16 | 0.01 | 0.04 | 0.06 | 0.06 | 0.08 | 0.09 |
| $H_2S$ | 7.31 | -0.03 | -0.04 | -0.06 | 0.07 | -0.08 | -0.09 |
| CO | 6.16 | -0.05 | -0.01 | 0.01 | 0.00 | 0.03 | 0.03 |
| $N_2$ | 5.12 | -0.01 | 0.03 | 0.06 | 0.04 | 0.07 | 0.08 |
| $CO_2$ | 5.60 | -0.01 | 0.03 | 0.05 | 0.05 | 0.07 | 0.07 |
| $CH_2O$ | 7.39 | -0.01 | -0.04 | -0.02 | 0.02 | 0.01 | 0.01 |
| MAE |  | 0.08 | 0.07 | 0.07 | 0.08 | 0.07 | 0.07 |
| MaxAE |  | 0.24 | 0.20 | 0.18 | 0.19 | 0.15 | 0.15 |

**Table 3:** Multicomponent DFT proton affinity errors relative to the experimental values, mean absolute errors (MAE), and maximum absolute errors (MaxAE) for different electronic basis sets. A negative proton affinity error indicates the multicomponent DFT proton affinity is smaller than the experimental value. All values are reported in eV.

It has previously been shown that the accuracy of multicomponent CC proton affinities has a strong dependence on the choice of electronic basis set.[16] Therefore, we also calculated multicomponent CCSD(T) protonic affinities using the same procedure as the multicomponent

DFT proton affinity calculations. All single-component CCSD(T) calculations in this study were performed using CFOUR.[52] All multicomponent CCSD(T) calculations were performed using our multicomponent CCSD(T) code that is available for free on GitHub.[53] The results from these calculations are presented in Table 4.

Unlike multicomponent DFT proton affinities, multicomponent aug-cc-pV$n$Z-mc CCSD(T) proton affinities are more accurate relative to aug-cc-pV$n$Z proton affinities at an identical zeta level. For a given zeta level, the MAEs of the aug-cc-pV$n$Z-mc electronic basis sets are similar in accuracy to the MAEs of the aug-cc-pV$n$Z electronic basis set one zeta level higher.

| Molecule | Experiment | aug-cc-pV$nz$ | | | aug-cc-pV$nz$-mc | |
|---|---|---|---|---|---|---|
| | | DZ | TZ | QZ | DZ | TZ |
| $CN^-$ | 15.31 | -0.56 | -0.23 | -0.11 | -0.28 | 0.01 |
| $NO_2^-$ | 14.75 | -0.42 | -0.15 | -0.05 | -0.11 | -0.17 |
| $NH_3$ | 8.85 | -0.34 | -0.14 | -0.02 | -0.08 | 0.04 |
| $HCOO^-$ | 14.97 | -0.45 | -0.17 | -0.05 | -0.13 | 0.07 |
| $HO^-$ | 16.95 | -0.49 | -0.19 | -0.05 | -0.17 | 0.03 |
| $HS^-$ | 15.31 | -0.56 | -0.24 | -0.11 | -0.24 | 0.00 |
| $H_2O$ | 7.16 | -0.40 | -0.16 | -0.05 | -0.14 | 0.01 |
| $H_2S$ | 7.31 | -0.33 | -0.11 | 0.01 | -0.04 | 0.10 |
| CO | 6.16 | -0.35 | -0.11 | -0.01 | -0.10 | 0.10 |
| $N_2$ | 5.12 | -0.37 | -0.13 | -0.03 | -0.13 | 0.05 |
| $CO_2$ | 5.60 | -0.36 | -0.15 | -0.06 | -0.09 | 0.04 |
| $CH_2O$ | 7.39 | -0.34 | -0.14 | -0.02 | -0.07 | 0.07 |
| MAE | | 0.41 | 0.16 | 0.05 | 0.13 | 0.06 |
| MaxAE | | 0.56 | 0.24 | 0.11 | 0.28 | 0.17 |

**Table 4:** Multicomponent CCSD(T) proton affinity errors relative to the experimental values, mean absolute errors (MAE), and maximum absolute errors (MaxAE) for different electronic basis sets. A negative proton affinity error indicates the multicomponent DFT proton affinity is smaller than the experimental value. All values are in eV.

The final benchmark is the lowest lying protonic excitation energies[11, 13, 19] of HCN and FHF$^-$. These excitation energies are a common benchmark for new excited-state multicomponent methods. In this study, the protonic excitation energies were computed with multicomponent HCI.[20-22]

Multicomponent HCI calculations are carried out in a two-stage procedure. In the first stage the variational energy is calculated using a selected configuration interaction procedure. The second stage corrects the energy using second-order Epstein-Nesbet perturbation theory.[54-55] The selection process is controlled by two user-selected parameters, $\varepsilon_{\text{VAR}}$ and $\varepsilon_{\text{PT2}}$, for the variational and perturbative stages, respectively. More detail about multicomponent HCI can be found in the literature.[20-22, 56-58]

We performed excited-state HCI calculations with mixed cc-pV$n$Z and cc-pV$n$Z-mc electronic basis sets. These calculations are labeled as XZ/YZ where X and Y are the zeta levels for the non-hydrogen atoms and hydrogen atom, respectively. All multicomponent calculations used the PB4F-2 protonic basis set, with geometries obtained from single-component CCSD optimizations using the aug-cc-pVTZ electronic basis set, which is identical to the procedure in the previous excited-state multicomponent HCI study.[21] For all calculations, the four lowest energy states were calculated with the smallest possible values of $\varepsilon_{\text{VAR}}$ and $\varepsilon_{\text{PT2}}$ given the limitations of the multicomponent HCI code and our computational resources. As was done previously,[21] we restricted the allowed electronic excitations to a maximum of triple excitations.

The results from the multicomponent HCI calculations are shown in Table 5. For reference, we performed excited-state FGH calculations on the HCN and FHF$^-$ molecules. In these FGH calculations, the electronic basis functions of the quantum proton were allowed to move, which, as discussed in the literature,[21] functionally increases the size of the electronic basis set associated with the quantum proton and makes a quantitative comparison between the multicomponent and FGH calculations difficult. However, the FGH energy can still be used to estimate the value of the protonic excitation energy when using a large electronic basis set centered on the quantum proton, a complete protonic basis set, and when all electron-nuclear correlation is included.

Compared to protonic densities and proton affinities, the cc-pV$n$Z-mc electronic basis set calculations show smaller improvements in the protonic excitation energies compared to cc-pV$n$Z electronic basis set calculations when using the FGH energies as a reference. As previously discussed,[21] this is, in part, due to fact that the FGH calculations were performed with an electronic basis set that is allowed to move with the hydrogen atom, which makes it difficult to compare to multicomponent calculations with smaller electronic basis sets. However, the DZ/DZ-mc basis set results are still more accurate than the DZ/DZ basis set results for both HCN and FHF$^-$ and the DZ/TZ-mc results are more accurate than the DZ/TZ results for HCN. The DZ/TZ results for FHF$^-$ are more accurate than the DZ/TZ-mc results for the bending mode. Some of the error in the protonic excitation energies is likely due to an incomplete description of electron-nuclear correlation, but given that all multicomponent protonic excitation energies remain far from the FGH TZ/TZ values along with the large change in the DZ/DZ-mc and DZ/TC-mc protonic excitation energies, it appears that large electronic basis sets are still required for excited-state calculations.

### HCN

| Elec. Basis | $\varepsilon_{VAR}$ | $\varepsilon_{PT2}$ | Var Bend | Var Stretch | Var + PT2 Bend | Var + PT2 Stretch |
|---|---|---|---|---|---|---|
| DZ/DZ-mc | 1.0E-05 | 6.0E-08 | 2224 | 3797 | 2217 | 3793 |
| DZ/TZ-mc | 3.0E-05 | 2.0E-08 | 1747 | 3678 | 1699 | 3659 |
| TZ/TZ-mc | 5.0E-05 | 5.0E-07 | 1920 | 3718 | 1704 | 3632 |
| DZ/DZ | 7.5E-06 | 5.0E-08 | 2973 | 4277 | 2968 | 4274 |
| DZ/TZ | 1.5E-05 | 7.0E-07 | 2165 | 3867 | 2147 | 3847 |
| FGH TZ/TZ | | | 685 | 3100 | | |

### FHF$^-$

| Basis | $\varepsilon_{VAR}$ | $\varepsilon_{PT2}$ | Var Bend | Var Stretch | Var + PT2 Bend | Var + PT2 Stretch |
|---|---|---|---|---|---|---|
| DZ/DZ-mc | 8.0E-06 | 3.0E-07 | 2379 | 3392 | 2378 | 3391 |
| DZ/TZ-mc | 3.0E-05 | 1.0E-07 | 2363 | 3114 | 2338 | 3082 |
| TZ/TZ-mc | 7.0E-05 | 5.0E-07 | 2074 | 2836 | 1859 | 2670 |
| DZ/DZ | 8.0E-06 | 3.0E-07 | 2772 | 3843 | 2771 | 3842 |
| DZ/TZ | 3.0E-05 | 4.0E-07 | 2311 | 3143 | 2251 | 3091 |
| FGH TZ/TZ | | | 1290 | 2044 | | |

**Table 5:** Multicomponent HCI protonic excitation energies relative to the ground states of HCN and FHF$^-$ with different electronic basis sets. Bend is the first excited state and is doubly degenerate. Stretch is the third excited state. More information about these definitions can be found in Reference [21]. All protonic excitation energies are in cm$^{-1}$. DZ/DZ, TZ/TZ, and FGH results are taken from Reference [21].

As mentioned in the Introduction, the ideal cc-pV*n*Z-mc electronic basis sets would calculate single-component energies identical to the cc-pV*n*Z electronic basis sets while improving calculated protonic properties. However, the addition of new electronic basis functions for hydrogen will almost always change the single-component energy, similarly to what happens when a mixed electronic basis set is used in multicomponent CCSD.[16]

To test how much the single-component energy changes, we performed single-component DFT and CCSD(T) calculations using the cc-pV*n*Z, cc-pV*n*Z-mc and mixed electronic basis sets on the same set of 12 small molecules for which the proton affinities were calculated. For the calculations on molecules with multiple hydrogen atoms, only the most acidic hydrogen atom used the cc-pV*n*Z-mc or larger mixed electronic basis set. The remaining hydrogen atoms used the cc-pV*n*Z or smaller mixed electronic basis set. All calculations were performed at the geometry of the cc-pV6Z optimized minimum for DFT and the geometry of the cc-pVQZ optimized minimum for CCSD(T). The mixed calculations are labeled using the same convention as the protonic excitation energies. Summary statistics from these calculations are given in Table 6. Complete results for all calculations can be found in the Supplemental Materials. From these calculations, we see that the new cc-pV*n*Z-mc electronic basis sets change the single-component electronic energy less than a mixed electronic basis set does, which indicates that the cc-pV*n*Z-mc electronic basis sets can be used with less concern that their accuracy may be due to a cancelation of errors.

|         | cc-pV*n*Z |       |       | cc-pV*n*Z-mc |       |        |
|---------|-----------|-------|-------|--------------|-------|--------|
|         | DZ/TZ     | DZ/QZ | TZ/QZ | DZ           | TZ    | QZ     |
| DFT     | 5.11      | 6.06  | 0.583 | 2.31         | 0.205 | 0.0503 |
| CCSD(T) | 9.42      | 16.3  | 3.77  | 2.78         | 1.16  | 0.446  |

**Table 6:** MAD of the difference in total energy for 12 small molecules between DFT and CCSD(T) single-component calculations with the cc-pV*n*Z electronic basis set and either a mixed electronic basis set or the cc-pV*n*Z-mc electronic basis set. All values are in mHa. The mixed electronic basis sets are compared to cc-pV*n*Z calculations with a value of n equal to the n used for the non-hydrogen atoms.

In Table 7, we show the number of electronic basis functions for two of the systems in this study for the cc-pV*n*Z and cc-pV*n*Z-mc electronic basis sets. For systems with hydrogen atoms, the new cc-pVnZ-mc electronic basis sets contain more electronic basis functions than the cc-pVnZ electronic basis for a given zeta level. However, given the results presented in this section, it appears that calculations performed with the cc-pVnZ-mc electronic basis sets yield protonic properties as accurate as do calculations performed with cc-pVnZ basis sets of one or two zeta levels higher, for proton affinities and protonic densities, respectively. In the former case, this results in a reduction in the number of electronic basis functions by at least 35% for the HCN and HCOOH molecules. In the latter case, this results in a reduction in the number of electronic basis functions by at least a factor of two. Multicomponent methods have the same computational scaling as do their single-component counterparts, with the computational bottleneck normally being the solution of the purely single-component equations because they have more electronic orbitals than nuclear orbitals for most systems. Therefore, the cc-pVnZ-mc electronic basis sets should increase the computational efficiency of multicomponent calculations and allow for calculations on larger systems than calculations with the cc-pVnZ electronic basis sets.

|    | H | | HCN | | HCOOH | |
|----|---|---|---|---|---|---|
|    | cc-pVnZ | cc-pVnZ-mc | cc-pVnZ | cc-pVnZ-mc | cc-pVnZ | cc-pVnZ-mc |
| DZ | 5 | 10 | 33 | 38 | 52 | 62 |
| TZ | 14 | 28 | 74 | 88 | 118 | 146 |
| QZ | 30 | 59 | 140 | 169 | 225 | 283 |
| 5Z | 55 |  | 237 |  | 383 |  |
| 6Z | 91 |  | 371 |  | 602 |  |

**Table 7:** Number of electronic basis functions for three different systems with different electronic basis sets. For HCOOH, both hydrogen atoms use an identical electronic basis set. For the 6z basis set, *i* functions are included for non-hydrogen atoms, which differs with the fitting procedure for the exponents in Section II.

## IV. Conclusions

A new family of hydrogen-atom correlation-consistent electronic basis sets designed for multicomponent method calculations, cc-pV*n*Z-mc, has been introduced. The cc-pV*n*Z-mc electronic basis sets are shown to improve multicomponent DFT protonic densities and multicomponent CCSD(T) proton affinities compared to calculations with the cc-pV*n*Z electronic basis sets. The cc-pV*n*Z-mc electronic basis sets should increase the computational efficiency of many multicomponent calculations, which will allow multicomponent calculations to be performed on larger systems using multicomponent many-body methods or for longer time scales for multicomponent DFT *ab initio* molecular dynamics.

**Supplemental Materials**

Single-component DFT and CCSD(T) energetic difference between the cc-pV*n*Z-mc and cc-pV*n*Z electronic basis sets for all twelve small molecules are available free of charge.

**Acknowledgement**

KRB thanks the University of Missouri-Columbia for startup funding.

**Supplemental Materials**

**Table S1:** Difference in energy for 12 small molecules between DFT and CCSD(T) calculations with the cc-pV*n*Z electronic basis set and either a mixed electronic basis set or the cc-pV*n*Z-mc electronic basis set. All values are in mHa. The mixed electronic basis sets are compared to cc-pV*n*Z calculations with a value of *n* equal to the *n* used for the non-hydrogen atoms. More details about the calculation can be find in the main text.

| | DFT | | | | | |
|---|---|---|---|---|---|---|
| | cc-pV*n*Z | | | cc-pV*n*Z-mc | | |
| Molecule | DZ/TZ | DZ/QZ | TZ/QZ | DZ | TZ | QZ |
| HCN | 4.66E-03 | 5.58E-03 | 5.40E-04 | 2.56E-03 | 2.07E-04 | 5.76E-05 |
| $HNO_2$ | 7.83E-03 | 9.62E-03 | 1.08E-03 | 3.08E-03 | 3.60E-04 | 5.96E-05 |
| $NH_4$ | 5.07E-03 | 5.46E-03 | 3.67E-04 | 2.71E-03 | 1.72E-04 | 4.66E-05 |
| HCO | 4.71E-03 | 5.92E-03 | 4.52E-04 | 2.21E-03 | 1.96E-04 | 5.73E-05 |
| $H_2O$ | 7.96E-03 | 9.95E-03 | 1.10E-03 | 2.96E-03 | 3.39E-04 | 5.29E-05 |
| $H_2S$ | 2.48E-03 | 3.04E-03 | 3.61E-04 | 9.76E-04 | 1.20E-04 | 2.83E-05 |
| $H_3O$ | 6.15E-03 | 6.93E-03 | 5.77E-04 | 2.98E-03 | 1.95E-04 | 5.13E-05 |
| $H_3S$ | 2.23E-03 | 2.75E-03 | 3.46E-04 | 8.37E-04 | 1.11E-04 | 2.91E-05 |
| HCOOH | 6.13E-03 | 7.23E-03 | 7.50E-04 | 2.90E-03 | 2.47E-04 | 5.75E-05 |
| $HN_2$ | 4.22E-03 | 4.59E-03 | 4.21E-04 | 1.70E-03 | 1.65E-04 | 5.54E-05 |
| $HCO_2$ | 5.64E-03 | 6.52E-03 | 6.04E-04 | 2.55E-03 | 1.86E-04 | 5.34E-05 |
| $CH_2OH$ | 4.22E-03 | 5.16E-03 | 3.94E-04 | 2.28E-03 | 1.56E-04 | 5.51E-05 |
| MAD | 5.11E-03 | 6.06E-03 | 5.83E-04 | 2.31E-03 | 2.05E-04 | 5.03E-05 |

| | CCSD(T) | | | | | |
|---|---|---|---|---|---|---|
| | cc-pV*n*Z | | | cc-pV*n*Z-mc | | |
| Molecule | DZ/TZ | DZ/QZ | TZ/QZ | DZ | TZ | QZ |
| HCN | 7.92E-03 | 1.38E-02 | 3.10E-03 | 2.88E-03 | 9.30E-04 | 4.86E-04 |
| $HNO_2$ | 1.38E-02 | 2.37E-02 | 5.49E-03 | 4.06E-03 | 1.80E-03 | 5.64E-04 |
| $NH_4$ | 7.01E-03 | 1.14E-02 | 2.84E-03 | 2.40E-03 | 8.58E-04 | 4.22E-04 |
| HCO | 7.27E-03 | 1.30E-02 | 2.65E-03 | 2.34E-03 | 8.26E-04 | 4.40E-04 |
| $H_2O$ | 1.44E-02 | 2.44E-02 | 5.33E-03 | 4.03E-03 | 1.75E-03 | 5.42E-04 |
| $H_2S$ | 9.23E-03 | 1.69E-02 | 3.86E-03 | 1.77E-03 | 1.16E-03 | 3.48E-04 |
| $H_3O$ | 1.01E-02 | 1.65E-02 | 3.68E-03 | 3.18E-03 | 1.14E-03 | 4.43E-04 |
| $H_3S$ | 7.81E-03 | 1.43E-02 | 3.42E-03 | 1.53E-03 | 1.01E-03 | 3.22E-04 |
| HCOOH | 1.09E-02 | 1.81E-02 | 4.58E-03 | 3.46E-03 | 1.44E-03 | 5.30E-04 |
| $HN_2$ | 7.36E-03 | 1.33E-02 | 3.69E-03 | 2.11E-03 | 1.01E-03 | 3.88E-04 |
| $HCO_2$ | 9.49E-03 | 1.58E-02 | 3.82E-03 | 2.95E-03 | 1.20E-03 | 4.47E-04 |
| $CH_2OH$ | 7.84E-03 | 1.38E-02 | 2.83E-03 | 2.68E-03 | 8.28E-04 | 4.25E-04 |
| MAD | 9.42E-03 | 1.63E-02 | 3.77E-03 | 2.78E-03 | 1.16E-03 | 4.46E-04 |